\documentstyle[12pt]{article}
\begin{document}
\begin{center}
{\bf ON MESOATOMS OF THE DEUTERIUM AND POSSIBLE EXISTENCE OF EXOTIC
DIBARYONS}\\
\vspace*{1cm}

S.B. GERASIMOV\\
{\it Bogoliubov Laboratory of Theoretical Physics,\\}
{\it Joint Institute for Nuclear Research, Dubna, 141980 Russia}
\end{center}

\vspace*{.5cm}
\begin{abstract}
{\small
As a desirable supplement to the reaction $pp \to pp2\gamma$, proposed earlier  (nucl-th/9712064 and references therein) to
probe for the NN--decoupled dibaryon resonances we suggest
to use for the same goal the radiative capture
processes in mesoatoms of the deuterium which we
consider to be especially feasible for further test and invesigation of
possible low-lying exotic resonance states.}
\end{abstract}
\vspace*{1cm}

{\bf 1.}The experimental discovery of dinucleon (or, generally, multibaryon)
resonances not decaying into two (or more) nucleons in the ground state
would be one of the most spectacular explications of nonpotential, nonnucleon
degrees of freedom and would imply important consequences
in further development of nuclear and hadron matter physics.
The nonstrange NN-decoupled dibaryons with small widths appear to be the most
promising and interesting candidates for experimental searches. Among
available candidates to be confirmed (or rejected) in future dedicated and,
hopefully, more sensitive experiments we wish to mention the indications
of the existence of a narrow $d^{\prime}$ dibaryon slightly above the $\pi NN$
threshold, coming from data on pion double charge exchange (DCE) on nuclei
\cite{a4}, and $d_{1}$-enhancement below $\pi NN$-threshold, seen in
preliminary data on the proton-proton double bremsstrahlung reaction
at $200 MeV$ \cite {a3}.\\
It would be of undoubted interest to try different reactions to search
for these states. With the pion probe, the presumed $d_{1}$-dibaryon
can be excited via strong interactions only  on the three- (or more) nucleon
states. With the deuteron targets, most thoroughly investigated and most
easy to deal with theoretically, we have to resort to radiative processes.\\
This report aims at drawing attention to the real feasibility and suitability
of radiative $\pi (\mu)$-meson capture processes in the deuterium
mesoatoms for the sake of inquiry on possible nucleon and dinucleon exotics.
In what follows we concentrate mainly on the radiative pion capture
processes because they are experimentally easier to investigate.
This proposal is not entirely new. In fact, there is the work devoted
to search for the exotic isotensor ( with the isospin $I=2$) dibaryon
resonance done at TRIUMF \cite{a5}. We, however, propose to look for a rather
specific object, which seemed to be beyond the design and kinematics area
covered by the abovementioned experiment. Furthermore we relay on what
can be taken into account from an analysis \cite{a6} of situation
connected with the double bremsstrahlung experiment \cite{a3}.\\
A few remarks are to be made about quantum numbers of the mentioned
candidates.  The isospin 0 assignment for the $d'$ resonance with quantum
numbers $J^P=0^{-}$ was motivated by calculations within a QCD string
model, while the isospin $2$ assignment was made in \cite{a7} on the basis
of the $\pi NN$ bound system molel and within the Skyrme model approach.  With
the $d'$ quantum numbers $0^{-}$ and isospin 2, the dominant isobar in the
pion-nucleon sub-system is $P_{33}$ or $\Delta$, so that nucleon and
isobar have relative orbital angular momentum 1 and total spin 1.  With the
isospin 0 assumed, the dominant pion-nucleon sub-system in three-body model is
the $S_{11}$ isobar and the nucleon-isobar quasi-two-body
system has the relative orbital momentum and total spin both equl to 0.\\
For
the $d_1$-state, presumably seen in the proton-proton double bremsstrahlung
reaction below the pion threshold, we suggest the $\Delta N$ with relative
orbital momentum 0 as the dominant cluster configuration.  Of the possible
values 1 or 2  for spin and isospin, we consider the unit spin/isospin value
as more natural for the state with lowest mass.  In this case the NN decay
channel is strictly forbidden by the exclusion principle, and if the $\pi N$
decay mode is kinematically forbidden we have the radiative decay as the only
possible one with the width of $\sim$ KeV scale.  It might, however, be that
the isospin 2 assighment for $d_1$ is dynamically preferable, like in the
case of the $d'$-state mentioned above.  In that case the $d_1 \to NN $ decay
mode is possible though strongly suppressed being of order $O(\alpha ^2)$
versus the radiative decay of the order $O(\alpha)$. To discriminate between
two options $(I=0$ vs $2)$, it is important to study several reations
differently sensitive to different isospin values. It is with this feature in
view, that we propose to make use of the double radiative capture process  in
the pionic deuterium in addition to the nucleon-nucleon double bremsstrahlung
reactions. As will be seen in the next section, the transition operator
between the deuteron and dibaryon resonance state vectors is transformed, by
construction within the assumed model, as the isovector under rotations in the
isospin space.  This means that, unlike the proton-proton double
bremsstrahlung reaction, the narrow dibaryon will not be excited in the
radiative pion capture on the isosinglet deuteron if the isospin of this
resonance equals 2.

{\bf 2.}
The branching ratios of different $\pi^{-}$ - capture
channels in the pionic deuterium: $\pi^{-}d \to nn(.7375 \pm .0027), nn\gamma
(.2606 \pm .0027), nne^{+}e^{-} ((1,81 \pm .02)\cdot 10^{-3}), nn\pi^{0}((1.45
\pm .09)\cdot 10^{-4})$, measured at TRIUMF \cite{a8,a9}, give the
total probability $w(\pi^{-}d \to X(measured) = 1.000 \pm .038$. Therefore,
the upper bound of still undetected channels is not much larger than $\simeq
.38 \%$, well within capability of measurements if there is good signature .
We estimate which could be the yield of two $\gamma's $ from consequent
processes of the radiative excitation and de-excitation of the exotic
$d_{1}(1920)$, presumably seen in the proton-proton double bremsstrahlung:
\begin{eqnarray}
(\pi^{-}d)_{atom} \to \gamma d_{1} \to  \gamma \gamma X
\end{eqnarray}
The radiative decay branching ratio $Br(d_1 \to \gamma X)=1$, hence we
need to calculate only the $d_1$ - excitation probability, i.e. the
transition $(\pi_{-}d)_{atom} \to \gamma d_{1}(1920)$. The radiated
photon takes off the energy $\omega = 92.9 MeV$ thus enabling the resonance
state to be on its mass-shell.As a hint for possible qualitative estimate
we note that the probability of ordinary radiationless $\pi^{-}$-capture
$w((\pi^{-}d)_{atom} \to nn)$ is only three times as large as
that of the radiation capture.This is indication of the dominantly short-range
nucleon-nucleon interaction dynamics involved in a pure strong capture channel,
resulting in a poor overlap with the deuteron wave function having
characteristically large spatial extensions. An essential feature of
mechanisms of both the ordinary $NN$-channel \cite{a10} and
the assumed $d_{1}$-excitation is the appearance of the
$N \Delta$-configuration in an intermediate state of reactions considered.
So, it seems reasonable to expect that
\begin{eqnarray}
BR((\pi^{-}d)_{atom} \to \gamma d_1) \sim \alpha_{em} BR((\pi^{-}d)_{atom}
\to nn) \simeq .74/137 \simeq .5 \% .
\end{eqnarray}

More quantitative
estimation is made within a model used previously \cite{a6} for the reaction
$pp \to \gamma d_1$.  Namely, we assume the reaction mechanism when $\pi^{-}$
is radiatively captured by a nucleon to form the (virtual) $\Delta$ that, in
turn, is associated with a spectator nucleon to form the
$d_1(1920)$-resonance.\\
We adopt the explicitly phenomenological approach in our estimations.
Having in mind the completeness
of colourless hadron states,we shall estimate the probability of the
radiative transition $(\pi^{-}d)_{atom} \to \gamma d_{1}(IJ^{P}=11^{+} $
as a two-step
process,where the presumably lowest $NN$-decoupled state with the $J^P=1^{+}$,
we are tentatively calling $d_{1}$, is coupled with the initial or final
hadron states through the intermediate $N \Delta$-state with the same quantum
numbers.The $"\Delta"$ - symbol may also be referred to the virtual $\pi
N$-complex with quantum numbers of the $\Delta(1232)$-resonance but with a
different invariant mass.
We make use of standard formulas for a capture from atomic states following
from the assumption that the hadronic reaction is much shorter in range
than the atomic orbit radii. The rate is, schematically,
\begin{eqnarray}
w(L=0&\mbox{atomic state}) \sim |\psi(0)|^2 |\langle f|T_{\pi\gamma}(\vec{0})|i\rangle |^2
\end{eqnarray}

where $\psi(r)$ is the $L=0$ pionic atom wave function,
$\langle f|T_{\pi\gamma}(\vec{q})|i\rangle $ is the amplitude of the
reaction $\pi(\vec {q}) + N \to \gamma + \Delta$ with the free plane wave
of a pion with momentum $\vec{q} \simeq 0$, taken between the initial
$NN$-bound state (i.e. the deuteron) and the final $\Delta N$-state (i.e. the
$d_1$- resonance). As we deal with the threshold-type process, we proceed with
keeping only the seagull Feynman graph, approximating the $N\pi \gamma \Delta$
block ( the corresponding $N\gamma \pi N$-graph is known to give the
low-energy Kroll-Ruderman threshold theorem for the charged pion
photoproduction on nucleons). The $\Delta N\pi$-coupling constant is
defined by the $SU(6)$ symmetry through the known pion-nucleon coupling.
It seems justified also to neglect the retardation corrections,
i.e. we are using the long-wave approximation for the matrix element of
electric-dipole radiative transition.
The $d_{1}\Delta N$-vertex is described by a simple
form of the quasi-two-body wave function, for which the Hulthen-type radial
dependence was chosen by analogy with the deuteron radial wave function:
\begin{eqnarray}
R(r) = N\frac{1}{r}exp(-\alpha r)(1 - exp(-\beta(r-r_c)))
\end{eqnarray}

where N is the normalization
constant,$\alpha=\sqrt{2M_{red}\varepsilon_1},
\varepsilon_1=M+M_{\Delta}-M_{d_1},
M_{red}^{-1}=M^{-1}+M_{\Delta}^{-1},\beta=5.4 fm^{-1}, r_{c}=.5 fm$ and
$R(r)=0$ for $r \leq r_{c}$ is understood. The second factor in Eq.(2),
describing the behavior of wave function in the "interior" region
outside the hard core with the radius of $r_c=.5 fm$ is taken the same as in
the deuteron case.The radial part of the deuteron wave function is obtained
with $\varepsilon_1 \rightarrow \varepsilon = 2.23 MeV$
Taking the measured value of the total width $\Gamma_{tot} \simeq 1$ eV \cite
{a11} for the $1S$-level of pionic deuterium, we obtain the following
estimation for the (unobserved) decay channel
\begin{eqnarray}
BR((\pi^{-}p)_{atom} \rightarrow \gamma d_1 \to \gamma \gamma X)_{1S-state} =
.6 \%,
\end{eqnarray}
surpisingly close to scale estimate (2) and not
embarrasingly distant from the experimental bound $\le .4 \%$,
despite the adopted approximations being crude.
We note also, that our result does not
depend strongly on variation of the "effective" mass $M_{\Delta}$ from
$M+m_{\pi}$ to $M_{\Delta}=1232 MeV$, ($M=939 MeV, m_{\pi}=139 MeV$ being
masses of the nucleon and pion), if $M_{d_{1}}=1920 MeV$ is taken for granted.
To get an estimate of the background non-resonance $2\gamma$- emission rate,
we take
\begin{eqnarray}
BR(\pi^{-}d \to 2\gamma/1\gamma) \simeq BR(\pi^{-}p \to 2\gamma/1\gamma)
\simeq 1.3 \times 10^{-4},
\end{eqnarray}

where the corresponding ratio for the pionic hydrogen was calculated
by Beder \cite{a12}.\\
We have then
\begin{eqnarray}
BR((\pi^{-}d)_{atom} \to \gamma \gamma X)_{nonres} \simeq
BR((\pi^{-}d)_{atom} \to \gamma nn) \times 1.3 \cdot 10^{-4}
\simeq 3.4 \cdot 10^{-5}
\end{eqnarray}

which is considerably lower than the estimated resonance contribution.
We point out also a qualitative difference of the $\gamma_1 -
\gamma_2$ opening angle $\theta_{12}$ distribution following from the
resonance and nonresonance mechanisms. In the resonance excitation mechanism,
we have emission of the electric-dipole photon at the $d_1$-resonance
excitation vertex and the magnetic-dipole photon emission in the $d_1 \to
\gamma nn$ transition, the nn-pair being mainly in the $^{1}S_{0}$-state. The
polarization structure of the matrix element
\begin{eqnarray}
T(\vec{\epsilon_d}, \vec{\varepsilon_1} (k_1), \vec{\varepsilon_2} (k_2), ...)
\sim \ a_1 ([\vec{\epsilon_d} \times
\vec{\varepsilon_1}] \cdot [\vec{\varepsilon_2} \times \vec{k_2}]) +
(1 \leftrightarrow 2)
\end{eqnarray}

gives after the squaring and summation over polarizations
\begin{eqnarray}
W(\theta_{12},\varphi) = \frac{3}{16 \pi}\cdot (1+ \frac{1}{2} \cdot
\sin^{2}\theta_{12})
\end{eqnarray}

which has a maximum at $\theta_{12}=90^{o}$, while the corresponding
distribution in
the $\pi^{-}p \to 2\gamma n$ reaction, calculated by Beder \cite{a12},
and, by our assumption based on the impulse approximation,
in the $(\pi^{-}d)_{nonres} \to \gamma \gamma X$ -reaction, shows a shallow
minimum at $\theta_{12}=90^{o}$.\\

It can be noted in this respect that a recent
calculation \cite{a13} of the $\theta_{12}$ -distribution in reactions
$\pi^{-}A \to 2\gamma X$, $(A= ^{9}Be, ^{12}C) $  approximately
agrees with experiment \cite{a14,a15} for angles larger than $90^{o}$ but for
$\theta_{12} \le 90^{o}$ the calculations are consistently lower than data.
A possible role of the exotic resonance excitation is suggestive here, but
for a more quantitative estimation one has to take into account a number of
very important many-body effects: the Pauli blocking, Fermi-motion smearing,
collision broadening of the resonance propagating in nuclear matter. Indeed,
each inelastic $d_1N$-collision can trasform the "$\Delta$-part" of the
resonance into a nucleon via isovector, spin-dependent forces
trasmitted by pi- and rho-mesons, thus giving rise to a new decay channel
$d_{1}N \to 3N$.  Qualitatively, instead of
$\Gamma^{free}_{tot}(d_1)=\Gamma_{rad}(d_1) \simeq .5$ KeV we are led to use
$\Gamma^{matter}_{tot}(d_1) \simeq \rho \cdot v \cdot \sigma_{inel}(d_1N)$,
and for $\rho \simeq .17 fm^{-3}$, $v  =.2$ and $\sigma_{inel} \simeq 1$ mb,
we get $\Gamma^{matter}_{tot}(d_1)$ enhanced by 3 orders of magnitude
as compared to
$\Gamma^{free}_{tot}$.  That leads to $BR(\pi^{-}A \to 2\gamma X)$ of the
order $\le 10^{-5}$ in accord with measurements \cite{a14,a15}.  To
conclude this section, we are tempted to mention that the $\gamma$-spectra in
radiative pionic deuterium decays can, in principle, test narrow
baryon exotics of the type claimed in ref. \cite{a16}, where the evidence was
presented for three narrow baryon states with masses 1004, 1044 and 1094 MeV.
The first two of them are within reach of pionic mesoatoms studies.\\

{\bf 3.} Our conclusion is: In addition to the planned \cite{a17}
measurements of processes $\pi^{-}p \to 2\gamma n $ and double photon
capture on complex nuclei \cite{a18}, the radiative capture processes in
pionic deuterium as well as the continuum pion energy reactions
$\pi^{\pm}d \to \gamma (2\gamma) NN$ well deserve a devoted study
being a perspective source of potentially very important information.\\

This work was supported in part by
the Russian Foundation for Basic Research, grants No. 96-15-96423
and 96-02-19147.

\end{document}